\documentclass{iopart}

\usepackage{graphics}
\usepackage{epsf}
\usepackage{epsfig}

\begin{document} 


\title[Binder cumulant for Ising models]
{Critical Binder cumulant for isotropic Ising models on square and
  triangular lattices}

\author{W. Selke}

\address{
Institut f\"ur Theoretische Physik, 
RWTH Aachen,\\
52056 Aachen, Germany, \\ 
}

\ead{selke@physik.rwth-aachen.de}


\begin{abstract}
Using Monte Carlo techniques, the critical Binder cumulant $U^*$ of
isotropic nearest--neighbour Ising models on square and triangular
lattices is studied. For rectangular shapes, employing
periodic boundary conditions, $U^*$ is found to show the same
dependence on the aspect ratio for both lattice
types. Similarly, applying free boundary conditions for
systems with square as well as circular shapes for both lattices, the
simulational findings are also consistent with the suggestion that, for
isotropic Ising models with short--range interactions, $U^*$
depends on the shape and the boundary condition, but not on the
lattice structure.
 
\end{abstract}

\submitto{Journal of Statistical Mechanics: Theory and Experiment}


\section{Introduction}
The fourth order cumulant of the order parameter, the Binder
cumulant $U$ \cite{Binder1}, plays an important role in the theory
of phase transitions and critical phenomena. For 
instance, the transition point may be conveniently determined
from the intersection of the cumulant for different systems
sizes, and the universality class of the transition may
be identified from the critical exponent of the correlation
length which, in turn, can be obtained from the cumulant \cite{Binder1}.

The value of the cumulant at the transition point in
the thermodynamic limit, the critical Binder cumulant
$U^*$, has received much attention as
well \cite{Privman}, being a measure of the deviation of the 
distribution function for the order parameter from a Gaussian 
form. That value has sometimes believed to characterise
a given universality class, but conflicting findings show
that 'universality' holds in a rather restricted
sense. For concreteness and simplicity, we shall
consider here and in the following the universality class
of the two--dimensional Ising model. In that case, the critical
cumulant $U^*$ of the isotropic nearest--neighbour (nn) Ising model
on a square lattice with square shape, employing full periodic boundary
conditions, is known very accurately, from
transfer--matrix calculations augmented by
finite--size extrapolations, to
be $U^*= 0.61069...$ \cite{Kamin}. The same value seems to hold for
closely related two--dimensional isotropic nn Ising(--like) models,
including the nn 'border $\Phi ^4$ model' with softened spins,
the nn spin--1 Ising model, and XY as well as Heisenberg models
with an easy axis \cite{Kamin,Bruce,Janke,Holm,Schmid,Patry,Holtsch}.

Note, however, that $U^*$ may depend
on various details of the model, which do not affect the
universality class, in particular, the boundary
condition, the shape of the lattice, and the anisotropy of the
interactions \cite{Binder1,Privman,Kamin,Chen,Shchur,Dohm,Selke}.

In fact, it seems tempting to transcribe the dependence of $U^*$ on
the anisotropy of interactions for systems with fixed shape
into a shape dependence for the
isotropic case. Such mappings have been suggested
and confirmed to hold for
nn square lattice Ising models with different
horizontal, $J_h$, and vertical, $J_v$, couplings \cite{Kamin,Shchur}
as well as for
square lattice Ising models having anisotropic diagonal next--nearest
neighbour (nnn) interactions \cite{Chen,Shchur,Dohm,Selke}. In the
case of anisotropic nn Ising models with square shape, it is possible
to relate $U^*(J_v/J_h)$ to the critical cumulant $U^*(r)$
of the isotropic nn Ising models with rectangular shape, i.e. with
aspect ratio $r$ \cite{Kamin,Shchur}. In  the case of
the anisotropic nnn model with square shape, the corresponding
mapping seems to require isotropic nn Ising models with
rhombus or parallelogram
shapes \cite{Shchur,Dohm,Selke}. In general, it remains to be seen whether
Wulf shapes \cite{Abraham} at criticality provide a clue
for obtaining a universal critical cumulant, with the shape having
been adjusted to the anisotropy of the interactions \cite{Selke}.

In this article, we shall deal with the issue, whether, for
isotropic nn Ising models, $U^*$ depends only on
the shape of the system and the boundary
condition, but not on the lattice
structure. In that context, we shall
consider square and triangular lattices, estimating
$U^*$ in large-scale Monte Carlo simulations, especially for square
and rectangular shapes, using periodic and free
boundary conditions. In the case of 
periodic boundary conditions, we shall compare our simulational
findings for the triangular lattice with previous, highly accurate 
results on $U^*$ for the square lattice \cite{Kamin}.

The paper is organized as follows: In the next section, the model and
the Monte Carlo method are introduced. Then, in
the main part, the simulational results
will be presented. Finally, the findings will be summarized briefly.

\section{Model and Method}
\label{sec:model_simulations}

We consider isotropic spin-1/2 Ising models on square and triangular lattices
with ferromagnetic nearest neighbour (nn) interactions. The
Hamiltonian reads

\begin{equation}
{\cal H} = -J \sum\limits_{(i,j)} S(i) S(j)
\end{equation}

\noindent
where $S(i)= \pm 1$ is the spin at site $i$, and $J > 0$ denotes
the ferromagnetic interaction between spins at neighbouring lattices sites $i$
and $j$. The sum is taken over such pairs of neighbouring
sites. In the simulations, systems with $L_1= sL$ rows
and $L$ sites per row are studied (see Fig. 1), $s$ being a rational number
determining the aspect ratio, as discussed below. Both full periodic
and free boundary conditions are employed.

\begin{figure}[t] 
\centering
\includegraphics[clip,width=9.0cm,angle=0]{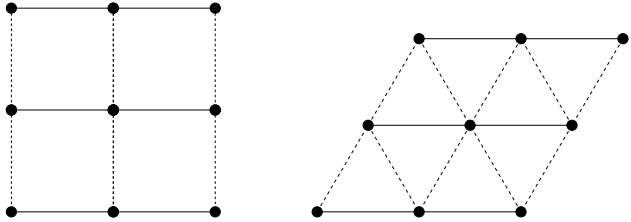}
\caption
{Sketch of square and triangular lattices. The rows are
denoted by solid lines
\label{fig:lattice_st}} 
\end{figure}

For both lattices, the
spin-spin correlations $<S_0S_r>$ are known to approach isotropy
for sufficiently large separation distance between spins
at site $0$ and $r$. 

Our aim is to study the Binder cumulant at the phase transition
temperature $T_c$. The exact critical temperature $T_c$
is known for both lattice types. For the square 
lattice, one has \cite{Onsager} 

\begin{equation}
k_{\mathrm B} T_{\mathrm c}/J= 2/\ln(\sqrt{2} +1)= 2.26918..\ldots \, .
\end{equation}

\noindent
For the triangular lattice, $T_c$ is
given by \cite{Houtappel}

\begin{equation}
k_{\mathrm B} T_{\mathrm c}/J=
2/\ln(\sqrt{3)} = 3.64095..\ldots\, .
\end{equation}

The Binder cumulant, the fourth order cumulant of the magnetization,
is defined, at fixed $s$, as \cite{Binder1}

\begin{equation}
U(T,L)
= 1 \, - \, \frac{\langle M^4\rangle}
{3 \langle M^2\rangle^2}
\label{eq:Binder_cum}
\end{equation}

\noindent
where $\langle M^2\rangle$ and $\langle M^4\rangle$ are the second
and fourth moments
of the magnetization, with the brackets $\langle ...\rangle$ denoting
thermal averages. For Ising models, the cumulant $U(T,L)$
approaches, in the thermodynamic
limit, the value 2/3 at temperatures $T < T_c$. It approaches the value
0 at $T > T_c$, reflecting the Gaussian
form of the magnetization distribution in the paramagnetic
phase \cite{Binder1}. At
$T_c$, in the thermodynamic limit, the critical Binder
cumulant $U^*$ acquires a nontrivial value, which will be estimated
for various cases in the following.

To obtain reliable estimates for $U^*$, we
performed large--scale Monte Carlo simulations \cite{Landau1}, using
single--spin--flip updates, being, supposedly, at least as
efficient as cluster--flip
algorithms for the system sizes we simulated (with up to about 7,000 sites).

Averaging always over several, rather long Monte
Carlo runs, error bars have been obtained in the standard
way by interpreting each realization as an independent event. We
did runs with up to $10^9$ Monte Carlo steps per site, for
each realization. Monitoring the
symmetry of magnetization histograms, the accuracy and reliablity of
the data may be checked. The simulations allowed
to estimate the critical cumulant $U^*$ to an accuracy of three
to four digits, see next section.

\begin{figure}[b] 
\centering
\includegraphics[clip,width=8.0cm,angle=0]{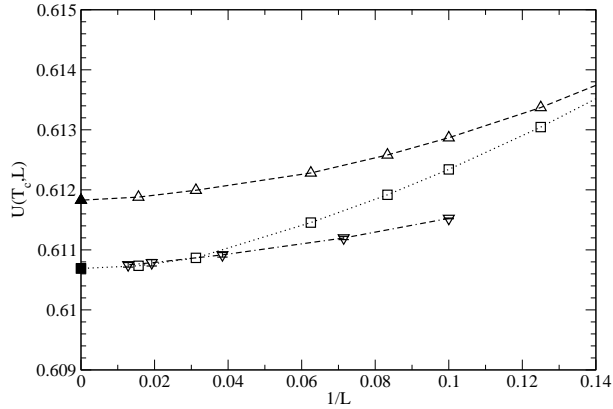}
\caption
{Size dependence of the Binder cumulant at $T_c$, $U(T_c,L)$, for
isotropic nn Ising
models on square lattices (squares) \cite{Selke} and triangular 
lattices (triangles
down) with square shapes (see text). For
comparison, the triangular case with rhombus
shape (triangles up) is included; here, as in the case of the
square lattice, error bars are smaller than symbol sizes
\cite{Selke}. In all cases, periodic boundary conditions
are employed. Lines are guides to the eye. Full symbols
refer to previous, highly accurate estimates of $U^*$ \cite{Kamin}.
\label{fig:bcu1}} 
\end{figure} 

We computed not only the cumulant, but also
other quantities like energy and specific heat, giving the possibility
for additional checks on the accuracy of the data.

\section{Results}
\label{sec:results}

We first consider the nn Ising model 
on the square lattice with square shape comprising $L_1= L$ rows
and $L$ sites per row, applying full periodic boundary
conditions. In that case, previous 
simulations \cite{Shchur,Selke}
showed that the critical Binder cumulant $U^*$
can be determined very accurately, when
smoothly extrapolating the data, obtained for system sizes up
to $L= 64$, to the thermodynamic limit. Indeed, the resulting estimate agrees
closely with that obtained from transfer--matrix
techniques augmented by systematic finite--size
extrapolations, $U^*= 0.61069...$ \cite{Kamin}, see Fig. 2.

\begin{figure}[b] 
\centering
\includegraphics[clip,width=8.0cm,angle=0]{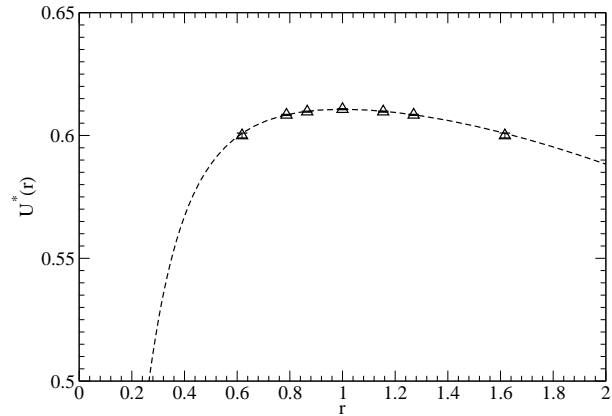}
\caption
{Simulational estimates of the critical cumulant $U^*(r)$ for
isotropic nn Ising models on triangular lattices (triangles) with
rectangular shapes of aspect ratio $r$, compared to previous
findings \cite{Kamin} for isotropic nn square lattice Ising
models with the same aspect ratio (dashed line). Periodic boundary
conditions are applied. 
\label{fig:aspectratio}} 
\end{figure} 

In the transfer--matrix study \cite{Kamin}, also
the isotropic nn square Ising model with rectangular shape (where
$s= L_1/L$ is now equal to the aspect ratio $r$) has been
analysed, using, again, full periodic boundary
conditions. The critical Binder cumulant, $U^*$, is observed
to depend continuously on $r$. $U^*(r)$ is maximal for
the square shape, $r=1$, and it is
obviously invariant under the transformation $r \longrightarrow
1/r$. The numerical findings $U^*(r)$ have been approximated very closely by
a polynomial expression of high order in
$4/(r+(1/r))-1$ \cite{Kamin}, $r$ being now a continuous
variable. That curve, $U^*(r)$, will play an
important role below, when interpreting some of the present
simulational data, see Fig. 3.

We now turn to isotropic nn triangular lattice Ising
models, employing again full periodic boundary conditons. The
systems have square and rectangular shapes, with $L_1= sL$ rows and
$L$ sites per row, as before. The aspect
ratio $r$ is now given by $r= s/g$, where the geometric factor $g$ 
for the triangular lattice may be easily seen to be $g= 2/\sqrt{3}$.

\begin{figure}[b] 
\centering
\includegraphics[clip,width=8.0cm,angle=0]{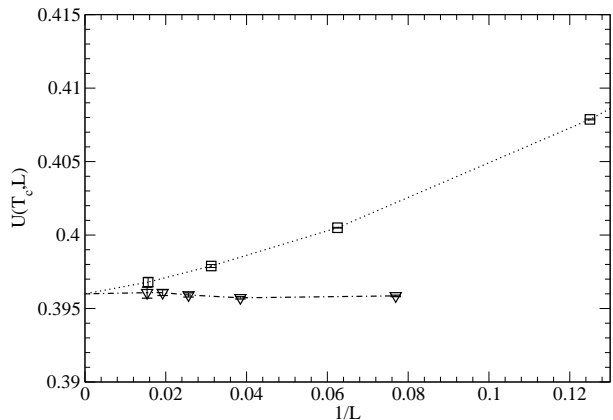}
\caption
{$U(T_c,L)$ vs. $1/L$ for isotropic nn Ising models on
square lattices (squares) and
triangular lattices (triangles) with square shapes (see text),
employing free boundary conditions. Lines are guides to the eye.
\label{fig:free boundary}} 
\end{figure} 

Obviously, the square shape, $r=1$, cannot be realized perfectly
for triangular lattices, because $s$ has to be an irrational
number, $s= g= 2/\sqrt{3}$. Nevertheless, for large lattices arbitrarily
close approximations may be obtained. In Fig. 2, $U(T_c,L)$ is shown, with 
$r=1.0104...$ for $L=14$ ($L_1=16$), and $r=1.0007..$ for
the larger lattices. As seen from Fig. 2, the resulting critical
Binder cumulant agrees very well with that of the isotropic nn square lattice
Ising model with square shape, $U^*= 0.61069...$. In contrast, the
isotropic nn triangular Ising model with rhombus shape (corresponding
to an anisotropic next--nearest neighbour Ising model
with square shape \cite{Shchur,Selke}), has a distinct, larger value
of the critical
Binder cumulant \cite{Kamin,Shchur,Selke}. For comparison, that case
is also displayed in Fig. 2, illustrating the shape
dependence of $U^*$ for isotropic models.  

Furthermore, one may analyse isotropic nn triangular Ising models
with rectangular shapes by fixing the ratio $s= L_1/L$, yielding
the aspect ratio $r= s/g$. From smooth finite-size extrapolations for
$U(T_c,L,r)$, similarly to those depicted in Fig. 2, one
may obtain accurate estimates for $U^*(r)$. Specificly, we studied
the cases $s=1.0, 1.1$, and 1.4, leading to irrational aspect
ratios, $r= s/g$. The results may then be
compared to the previous, highly accurate findings, discussed
above, on $U^*(r)$ for the
isotropic nn square Ising model with the same aspect ratios
\cite{Kamin}. The comparison is depicted in Fig. 3.

Fig. 3 presents clear evidence that identical critical Binder cumulants
hold for isotropic nn Ising models on square as well as on triangular
lattices, presuming the same aspect ratio of the
rectangular shapes and employing periodic boundary conditions.

In fact, this 'universal' aspect for isotropic models
seems to be true also for other boundary conditions and shapes. For
instance, in a recent Monte Carlo study \cite{Selke}, isotropic nn Ising
models on square and triangular lattices with circular shape
and free boundary conditions have been shown, to have, within
the accuracy of the estimates, the same value of
$U^*= 0.406 \pm 0.001$. This value has been suggested to
be the generic  value for two-dimensional
Ising models with Wulf shapes. Moreover, in the present study, we
simulated, employing free boundary conditions, isotropic nn
Ising models on square lattices with square shapes and on
triangular lattices with approximately square shapes (where $r=
1.0007...$ for all lattices sizes shown in Fig. 4). We obtain, for both
lattices, $U^*=0.396 \pm 0.001$, as seen from
Fig. 4. This observation is, again, consistent with that
universal aspect. Note that
$U^*$ takes in the case of the triangular lattice with
a rhombus shape, using free boundary conditions, a significantly
lower value, where 
$U^*= 0.379 \pm 0.001$ \cite{Selke}. This result reemphasizes the
shape dependence of the critical cumulant for isotropic models. 

In general, the irrelevance of the lattice type for $U^*$ in
case of isotropic Ising models with short--range interactions
may be understood \cite{Pokro} when expressing the
Binder cumulant in terms of two-- and four--spins correlation
functions, corresponding to the second and fourth
moments of the magnetization, respectively, see Eq. (4): With
these functions having, for isotropic models, a lattice independent
asymptotics for large separation distances, at criticality, $U^*$
may depend, indeed, only on the shape and the boundary condition, but
not on the lattice type.

\section{Summary}
\label{sec:sum}

The critical Binder cumulant $U^*$ for two--dimensional isotropic
nearest--neighbour Ising models
on square and triangular lattices has been
studied. From large--scale Monte Carlo simulations, accurate
estimates of $U^*$ have
been obtained from monitoring the size--dependence of the 
cumulant at the critical temperature.

Employing periodic boundary
conditions and considering rectangular shapes with the aspect
ratio $r$, a function $U^*(r)$, which had been determined
before for nn square lattice Ising models, is shown to hold for
isotropic nn Ising models on triangular lattices as well.

For free boundary conditions, we presented numerical evidence
that the same critical cumulant $U^*$ holds for 
isotropic nn Ising models on square as well as on triangular 
lattices, for fixed shape (circle and square).

Based on these selected examples, the simulational findings support the
suggestion that, for isotropic Ising models with short range
interactions, the critical Binder cumulant depends on shape and
boundary conditions, but not on the lattice structure
(and other irrelevant details of the interactions, not affecting the 
asymptotics of the isotropic spin--spin correlations ).  

Further studies on the interplay of anisotropy and
shape are encouraged. In particular, the possible universality of
$U^*$ for systems with Wulf shapes may be tested. Likewise, it
seems interesting to study in which way one
may transcribe the shape dependence
of $U^*$ for isotropic models into a dependence on the
interactions for anisotropic models with fixed shape.

I should like to thank H. Bl\"ote, V. Dohm,
M. Holtschneider, A. Hucht, S. L\"ubeck, V. L. Pokrovsky, and D. Stauffer
for useful discussions and information.

\section*{References}
\bibliographystyle{prsty}
\bibliography{./library}

\end{document}